\documentclass[reqno,dvips,12pt]{article} 
\usepackage[dvips]{graphicx}
\usepackage{amsmath,amssymb,amsfonts}
\newtheorem{theorem}{\qquad Theorem}
\newtheorem{proposition}{\qquad Proposition}

\newcommand{\al}{\alpha}

\newcommand{\wt}{\widetilde}
\newcommand{\cE}{{\cal{E}}}

\newcommand{\pa}{\partial}

\newcommand{\ve}{\varepsilon}
\newcommand{\be}{\beta}
\newcommand{\la}{\lambda}
\newcommand{\ka}{\kappa}

\newcommand{\nn}{\nonumber}

\begin{document}
\title{The law of large numbers for completely random behavior 
of market participants. Quantum economics}
\author {V.~P.~Maslov}
\date{}
\maketitle

\begin{abstract}
In this paper, we briefly discuss a mathematical concept 
that can be used in economics.
\end{abstract}


It has been known since Jacob Bernoulli that averaging in
economics is nonlinear and does not obey the rules of ordinary
arithmetics. For example, the average of winning or losing,
say, \$100,000 is not equivalent to customary life for a person
who is not wealthy, since losing could mean their becoming
homeless and even going to jail, whereas winning would not raise
their living standards high enough to justify the mere risk of
losing.  

Accumulating 51
interest. By pooling their capital, owners can become
monopolists and raise prices. This also corroborates 
the nonarithmetic nature of addition. Further, one observes
nonlinear addition when buying goods wholesale: the more you
buy, the less you pay per item.  

The mathematical problem of nonlinear averaging was attacked by
the outstanding Russian mathematician Kolmogorov, who obtained a
general formula for the average. I supplement his axioms by an
additional axiom saying that if a (small) number is added to
each term in the sum, then the average increases by same number.

This unambiguously results in the following rule for the average
of two numbers $a$ and $b$:   
$(a\text{ ``+'' }b)/2  = 1/\beta\log[(2^{\beta a}a + 2^{\beta b}b)/2]$, 
where $\beta$ is an unknown parameter and base~$2$ logarithm 
is used. For $\beta=0$, one arrives at the conventional linear
average. 
It is easily seen that $\beta$ is negative for purchase averaging 
and positive for sale averaging.  

If the operation 
$a\text{ `+' }b = 1/\beta\log [ 2^{\beta a} + 2^{\beta b}]$ is
taken as addition and the conventional addition  $a+b$ as
multiplication, 
then the commutativity, associativity, and distributivity laws 
hold. This arithmetics is possibly used somewhere in the
``Kingdom of Distorting Mirrors.''  

In my opinion, this arithmetics is more adequate to a market
economy than the conventional, classical arithmetics and permits
one, using model examples, to explain stock price breakout,
default, and a number of laws observed in statistical data
processing. 

In this note, I wish to present an economic effect that proved
quite unexpected to myself. I speak of a statistical sensation
that would be nicknamed ``fool's luck'' by ordinary people.
Since 2003, studies related to the analysis of operations at
London stock exchange have been evoking quite a response. About
five hundred papers that can be found in the Internet  under the
keywords zero intelligence discuss  the problem of completely
random behavior of market participants. The point is that
traders buying and selling stock have to take into account so
many various factors that they cannot make the right decision
unambiguously, and so the behavior of market participants does
not differ from a random behavior very much. And if it does, the
trader usually loses. Here we assume that the sales volume, the
number of participants, the number of nomenclatures of financial
tools for each price, and the number of tools are sufficiently
large. 

Thus what is a random choice?

To make a random choice, one should first calculate all possible
choices of purchases or sales and average them. The
parameter~$\beta$ can relatively easily be expressed via the
budget restraint (BR) in the first case and via the required
rate of return (RRR) in the second case. 

Note that possible choices of purchases and sales obey the
so-called Bose statistics. For example, the number of ways to
buy nails and bread for 2 cents is as follows: 
1) 2 cents for nails; 2) 2 cents for bread; 
3) 1 cent for nails and 1 cent for bread. 
All in all, there are three ways, since bank notes or
coins of the same value are indistinguishable.  
(The law that can be stated as ``money does not smell": nothing
changes if one note is exchanged for another.) 

How to make a random choice? It suffices to number all possible
purchases whose price does not exceed BR and then choose a
number randomly, as if throwing dice.  Computer generates random
(more precisely, ``pseudorandom") numbers. It turns out that the
law of large numbers holds: given a BR, a vast majority of such
random choices gives the same value for the number of purchases
at a given price. This value can be determined in a very simple
manner via the nonlinear average of all possible purchases by a
small variation of the given price.  (The number of tools
purchased at a given price is equal to the partial derivative
with respect to this price of the nonlinear average over all
possible purchases.)  A similar law holds for sellers. 

Note that the parameter $\beta<0$ for purchasers can also be
determined by equating the derivative of the nonlinear average
with respect to $1/\beta$ with the logarithm of the number of
possible purchases whose price does not exceed BR. Accordingly,
for sellers ($\beta>0$) the derivative should be equated 
with the logarithm of the number of possible sales that give a
profit not less than RRR. 

Thus if the behavior of market participants is completely
random, then we can find the number of goods purchased 
by a majority for a given price and even estimate the fraction
constituted by this majority. Conversely, if we cannot do that,
then the behavior of market participants cannot be viewed as
random.  

If the prices can be varied, then, by equating the amount of
goods sold as a function of RRR with the amount of goods
purchased as a function of BR, we can find equilibrium prices.  

It follows from an analysis of papers in the ``zero intelligence"
series that traders who do not subtilize, i.e., who act at
random rather than deliberately, mainly do not lose.  It is not
without reason that one speaks of fool's luck or beginner's
luck:  he who gambles for the first time (and hence has not yet
been spoiled by calculations) does not lose as a rule. There is
a large body of corroborating statistical evidence.  Hence if we
assume that a vast majority of traders have zero intelligence,
then the law established by the author gives a right forecast of
free market prices. 

Let us discuss a similar situation for consumer goods. Suppose
that a customer who has a certain amount BR of money for buying
gifts in advance for a large number of people enters a gift shop
(say, in Mexico) with $k$ shop floors each of which offers a large
variety of souvenirs for the same price.  The consumer does not
know the tastes of all acquaintances whom the gifts are intended
for and buys at random. Then the above-mentioned law applies,
and one can rather accurately predict how much money the
consumer will spend at each shop floor.  More precisely, if
there are many similar consumers, then a vast majority of them
will spend exactly the amount predicted by this law at each shop
floor. 

This is a very simple model, which can be generalized. For
example, consider the case in which the goods are divided
into~$i$ clusters with close prices within each cluster. 
Then one determines the (nonlinear) average profit over all
goods if  $N_i$ goods are sold for these prices and then the
average price for the $i$th cluster is found.  
The above-mentioned law remains valid in this case and
determines the amount spent by a customer for a given cluster of
goods.  

In this paper, we briefly discuss a mathematical concept 
that can be used in economics.

\section{Nonlinear averaging in the sense of Kolmogorov} 

A sequence of functions $M_n$ determines the {\it{regular\/}}
type of the average if the following (Kolmogorov) conditions are
satisfied: 

I. $M(x_1, x_2, \ldots, x_n)$ is a continuous and 
monotone function in each variable.
To be definite, we assume that $M$ increases in each variable. 

II. $M(x_1, x_2, \ldots, x_n)$ 
is a symmetric function\footnote{In our case, 
the symmetry follows from the Bose statistics
for bank notes.}. 

III. The average of identical numbers is equal to their common
value: $M(x, x, \ldots, x) =x$.

IV. A group of values can be replaced by their average
so that the common average does not change: 
$$
M(x_1,\ldots, x_m, y_1, \ldots, y_n) 
= M_{n+m}(x \ldots, x, y, \ldots, y_n),
$$ 
where $x = M(x_1, \ldots, x_n)$.

\begin{theorem}[Kolmogorov]
Under conditions {\rm I--IY},
the average $M(x_1, x_2, \ldots, x_n)$ takes the form
\begin{equation}
M(x_1, x_2, \ldots, x_n)= \psi \frac{\varphi(x_1)+\varphi(x_2)+
\ldots + \varphi(x_n)}{n}, \label{Kol}
\end{equation}
where $\varphi$ is a continuous strictly monotone function 
and $\psi$ is its inverse.
\end{theorem}

For the proof of the theorem, see~\cite{Kolmog1}.

\section{The main averaging axiom}
It is rather obvious for a stable system 
that the following axiom must hold.

If the same value $\omega$ is added to $x_k$,
then their average increases by this value~$\omega$.

Obviously, the nonlinear averaging of $x_i$ 
under normal conditions must also increase by this value.
We take this fact as {\bf Axiom~5}.

This axiom leads to a unique solution in the nonlinear case, 
i.e., 
the linear case (the arithmetic mean) naturally 
satisfies this axiom, as well as a unique 
(up to the same constant by which we can multiply 
all the incomes~$x_i$) nonlinear function.

In fact, the incomes~$x_i$ are calculated in some currency 
and, in general, must be multiplied by a quantity $\be$,
which is responsible for the purchasing power 
of this currency, so that this constant (the parameter~$\be$)
must {\it a priori\/} be contained in the definition 
of the income.  
Hence we can state that there exists a unique nonlinear function
that satisfies Axiom~5.

The function $f(x)$ has the form 
\begin{equation}
f(x)=C\exp(Dx)+B, \label{1kv}
\end{equation}
where $C,D\ne0$ and $B$ are numbers independent of~$x$.

\section{Semiring, an example of self-adjoint linear operators}
We consider the semiring generated by nonlinear averaging
and the space $L_2$ ranging in this semiring.

First, we consider a heat equation of the form
\begin{equation}
\frac{\pa u}{\pa t} = \frac h2 \frac{\pa^2 u}{\pa x^2}.
\label{2kv}
\end{equation}

Here $h$ is a small parameter, 
but we do not use its smallness now.

Equation (\ref{2kv}) is a linear equation.
As is known, this means that 
if $u_1$ and $u_2$ are its solutions, 
then the linear combination 
\begin{equation}
u= \la_1 u_1 + \la_2 u_2 \label{3kv}
\end{equation}
is also its solution. 
Here $\la_1$ and $\la_2$ are constants.

Now we perform the following change. We set
\begin{equation}
u=e^{-\frac wh}.\label{4kv}
\end{equation}
Then we obtain the following nonlinear equation 
for the unknown function $w(x,t)$:
\begin{equation}
\frac{\pa w}{\pa t} + \frac 12 \left (\frac{\pa w}{\pa x}
\right)^2 - \frac h2 \frac{\pa^2w}{\pa x^2} = 0. \label{5kv}
\end{equation}

This well-known equation is sometimes called 
the B\"urgers equation\footnote{The usual B\"urgers equation is
obtained  from this equation by differentiating with respect
to~$x$ and applying the change $v=\frac{\pa w}{\pa x}$.}.

The solution $u_1$ of Eq.~(\ref{2kv}) 
is associated with the solution
$w_1 =-h \ln u_1$ of Eq.~(\ref{5kv}), 
and the solution $u_2$ of Eq.~(\ref{2kv})
is associated with the solution 
$w_2 =-h \ln u_2$ of Eq.~(\ref{5kv}). 
The solution (\ref{3kv}) of Eq.~(\ref{2kv}) 
is associated with the solution 
$w= -h\ln(e^{-\frac{w_1+\mu_1}{h}} +e^{-\frac{w_2+\mu_2}{h}})$,  
where $\mu_i = -h \ln \la_i$, $i=1,2$.

This implies that Eq.~(\ref{5kv}) is a linear equation, 
but it is linear in a function space,
where the following operations were introduced:

the operation of taking the sum \ \ $a\oplus b = -h \ln(e^{\frac{-a}{h}} +
e^{\frac{-b}{h}})$;

and the operation of multiplication \ \ $a\odot \la= a+\la$.

In this case, the change $w=-h\ln u$ takes zero to infinity  
and the unity to zero.
Thus, $\infty$ is a generalized zero in this new space: 
$\O= \infty$, 
and the usual zero is a generalized unity:  
${\bf{1}}=0$. 
The function space,
where the operations $\oplus$ and $\odot$ are introduced,
with the associated zero $\O$ and the unity ${\bf{1}}$ 
is isomorphic to the usual function space 
with the usual multiplication and addition.

This can be interpreted in the following way:
somewhere on another planet,
the people are used to deal with precisely these operations
$\oplus$ and $\odot$,  
and then Eq.~(\ref{5kv}) is a linear equation from their
viewpoint. 

Everything written here is, of course, trivial,
and the people on our planet need not study 
new arithmetic operations, 
because, using a change of the function,   
one can pass from Eq.~(\ref{5kv}) to Eq.~(\ref{2kv}),
which is linear in the usual sense.
But it turns out that 
the ``Kingdom of distorting mirrors'' given by this semiring
is related to the ``capitalistic" economics.

In the function space ranging in the ring 
$a\oplus b=-h\ln(e^{-\frac ah} + e^{-\frac bh})$, 
$\la \odot b = \la + b$,
we introduce the inner product 
$$
(w_1, w_2) = -h \ln \int e^{\frac{w_1+w_2}{h}} dx.
$$
We show that the product in this space has 
the following bilinear properties:
$(a\oplus b, c) = (a,c)\oplus(b.c)$ and
$(\la \odot a,c)= \la \odot(a,c)$. 
Indeed, 
\begin{eqnarray}
&&(a\oplus b,c)= -h \ln \left ( \int \exp
\left(\frac{-(-h\ln(e^{\frac{-a}{h}} + e^{\frac{-b}{h}})+c)}{h}
\right) dx\right) = \nn \\
&&=-h \ln \left( \int (e^{\frac{-a}{h}} + e^{\frac{-b}{h}})
e^{\frac{-c}{h}} dx\right)= -h \ln\left( \int e^{-\frac{a+c}{h}}
dx + \int e^{-\frac{b+c}{h}} dx \right)=(a,c)\oplus(b,c), \\ \nn
&&(\la\odot a,c) = -h\ln\int e^{-\frac{a+\la}{h}} e^{-\frac ch}dx
=\\ \nn
&& = -h \ln\left(e^{-\frac \la h} \int e^{-\frac{a+c}{h}}
dx\right)=\la+\ln\int e^{-\frac{a+c}{h}} dx= \la\odot(a,c).
\end{eqnarray}
We consider an example of self-adjoint operators 
in this space,
namely, the operator
$$
L:W \longrightarrow W\odot(-h\ln\left(\frac{(W')^2}{h^2} -
\frac{W''}{h}\right).
$$
And now we verify whether it is self-adjoint:
\begin{eqnarray}
&&(W_1,LW_2) =-h\ln \int e^{-\frac{W_1+LW_2}{h}} dx= \\ \nn
&&=-h\ln\int \exp \left[-\left(W_1+W_2-
h\ln\left(\frac{W'_2)^2}{h^2} - \frac{W''_2}{h}\right)\right)
/h\right] dx= \\ \nn &&= -h \ln\int e^{\frac{-W_1}{h}}
e^{\frac{-W_2}{h}}\left( \frac{(W_2')^2}{h^2} -
\frac{W''_2}{h}\right) dx = -h \ln\int e^{\frac{-W_1}{h}}
\frac{d^2}{dx^2} e^{\frac{-W_2}{h}} dx=
\\ \nn
&&=-h\ln\int\frac{d^2}{dx^2}
e^{\frac{-W_1}{h}}e^{\frac{-W_2}{h}}dx= -h\ln\int
e^{\frac{-W_1}{h}}\left(\frac{(W'_1)^2}{h^2}
-\frac{W''_1}{h}\right) e^{\frac{-W_2}{h}}dx= \\ \nn
&&=-h\ln\int\exp\left[-\left(W_1-h\ln\left(\frac{(W'_1)^2}{h^2}
-\frac{W'_2}{h}\right)\right) /h\right] dx= \\ \nn
&&= -h\ln\int
e^{-\frac{LW_1+W_2}{h}} dx =(LW_1,W_2).
\end{eqnarray}
Its linearity can also be verified easily.

We construct the resolvent operator of the B\"urgers equation:
$L:W_0\rightarrow W$, where $W$ is a solution of Eq.~(\ref{5kv})
satisfying the initial condition $W|_{t=0}=0$.

The solution of Eq.~(\ref{2kv})
satisfying the condition $u|_{t=0}=u_0$ 
has the form
$$
u=\frac{1}{\sqrt{2\pi h}} \int e^{-\frac{(x-\xi)^2}{2th}} u_0(\xi)
d\xi.
$$
Taking into account that $u=e^{-\frac Wh}$ and $W=-h\ln u$,
we obtain the resolvent $L_t$ of the B\"urgers equation
\begin{equation}
L_tW_0 =-\frac{h}{\sqrt{2\pi h}} \ln \int
e^{-\left(\frac{(x-\xi)^2}{2th} + \frac{w(\xi)}{h}\right)} d\xi.
\end{equation}
The operator $L_t$ is self-adjoint in the new inner product.

\section{Entropy for the producer and the consumer.
Condition for the producer income and the consumer expenditure.
Production and consumption.
Equilibrium prices}

We consider a large group of producers
manufacturing goods of $M$ types. 
The corresponding {\it production structure\/}
is characterized by the vector 
$\omega^N=\langle K_1,\dots, K_M\rangle$, 
where $K_i$ is the number of goods of the $i$th type,
$i=1,\dots, M$, and $\sum\limits_{i=1}^M K_i=N$. 
The {\it consumption structure\/} is treated 
similarly. 
Suppose that $\epsilon_i$ is the price of goods of the $i$th 
type.
The income obtained by selling $N$ units of goods
is equal to $E=\sum\limits_{i=1}^M \epsilon_i K_i$.

The concepts introduced below 
are based on the notion of nonlinear averaging  
of the incomes obtained by realizing $N$ units of goods  
over all possible versions $\omega^N$  
of the production structure
for a given positive value of the parameter~$\beta$:
\begin{eqnarray}
M_N= \frac{1}{\beta}\log\left(\frac{1}{L}
\sum\limits_{K_1+\dots+K_M=N}2^{\beta\sum\limits_{i=1}^n\epsilon_i
K_i}\right). \label{average-1}
\end{eqnarray}
The nonlinear averaging is discussed in detail in~\cite{Mas2003}.

\subsection{Psychology of an ordinary depositor.
Psychological law of status quo preservation}

We note that losing \$100,000  
is much heavier in its psychological ``cost" 
than winning the same sum.
This means that an ordinary person prefers 
to preserve status quo, 
i.e., not to take risk of losing \$100,000.  
Therefore, 
if a person deposits a certain sum $N$ in several banks
at high interest, he must calculate what sum is sufficient
for him to live like a rentier, to preserve status quo,
and, accordingly, how to spread the money over several banks 
so as not to take any risk and not to lose his status.

This purely psychological fact 
is the base of our mathematical calculations.
We note that an ordinary depositor can rather easily 
calculate the sum of income, but it is difficult for him 
to evaluate the reliability of a bank proposing 
high interest rates.  
Therefore, in the formulas given below, 
the free parameter~$\be$ can be determined 
in terms of the sum of income introduced above, 
and then we obtain graphs showing 
how the bank deposits $n_i$
depend on the {\it a priori\/} given income.

The problem of calculating the ``surviving probability," 
which is close to the ``status quo preservation law,''
has been discussed by specialists 
in mathematical economics.

If the required rate of return (RRR) 
$\sum n_i\la_i= E$ is assumed to be an independent variable,
then the nonlinear averaging based on the Kolmogorov axioms
and the additional axiom proposed by the author 
is unique.

We assume that production runs normally
if the producer's income $\sum_1^n \ve_i N_i$ 
is larger than or at least equal to a quantity~$E_1$.
The consumer cannot spend more that 
$E_2 =\sum_1^n  \ve_i N_2$ (the budget restraint -- BR).
We consider the ``entropies" of the producer 
and the consumer as follows:
$H_1(\ve_1,\cdots, \ve_n, E_1)$ is the base~$2$ logarithm
of the number of sale versions  
for a sum no less than $E_1$; \ 
$H_2(\ve'_1, \cdots, \ve'_n, E_2)$ is the base~$2$ logarithm 
of the number of purchase versions  
for a sum not exceeding~$E_2$.

Let $\tilde{M}$ be the sum (\ref{average-1}) 
for $\be <0$, and let $\theta=\frac1\be$.

The equation 
$$
\frac{\pa M}{\pa \theta} =H_1
$$
allows us to obtain $\theta =\theta_1(H_1)$, 
and the equation 
$$
\frac{\pa \tilde{M}}{\pa \theta} =H_2
$$
allows us to obtain $\theta =\theta_2(H_1)$.

We choose $k$ possible purchase versions at random.

It turns our that, for the majority of these versions,  
the sum of money spent for buying goods at price~$\ve_i$, 
is ``almost" equal to 
$$
\frac{\pa M}{\pa \ve_i}|_{\theta=\theta_1}
$$
(this is an analog of the law of large numbers; 
the exact estimates are the same as in the usual law 
of large numbers).

A similar statement holds for the seller. 
The equilibrium prices follows from the relation
$$
\frac{\pa M}{\pa \ve_i}|_{\theta=\theta_1(\ve)} 
= \frac{\pa \tilde{M}}{\pa \ve_i}|_{\theta=\theta_2(\ve)}, \ \ 
i=1,\cdots, n.
$$
Thus, the ``resources--producer--consumer--etc.'' 
vertical line is divided into pairs each of which 
contains two new numbers, RRR and BR. 
The entropy (the Kolmogorov complexity) and 
the degree of risk are ``hidden" in the intermediate calculations. 
This is how the usual model of economical equilibrium 
(the general equilibrium) varies.

The models of dynamical equilibrium 
(the intertemporal general equilibrium)
vary similarly according to formula (41) 
given in \cite{TeorVer2004}, p.~276.

The price equilibrium condition is determined by the relations
\begin{equation}
\frac{\pa M}{\pa \ve_i}|_{\theta=\theta_1(\ve,E_1)} 
= \frac{\pa tilde{M}}{\pa \ve_i}|_{\theta=\theta_2(\ve,E_2)}.
\label{6kv}
\end{equation}.

It follows from the ``pair'' law that 
the derivative of the average  
with respect to the ``temperature" $\theta$ 
is the entropy 
and the derivative with respect to the price 
is the quantity of goods.
Relation (\ref{6kv}) is well known in the linear case.
Here we generalize it to the nonlinear case. 
This generalization allows us to take 
the Kolmogorov complexity into account, 
and hence the entropy, 
which is one of the most important notions in economics.
Its conjugate, the temperature, determines the degree of risk,
and sometimes, the volatility. 
But the variables in the final formulas
contain only the incomes $RRR$ and the expenditures $BR$,
which can be calculated easily.
Thus,~$H$ and~$\theta$ are ``hidden parameters" here.

\section{Tunnel canonical operator in economics}

The equilibrium prices are determined by the condition
that the demand and offer must be the same 
for each item of goods and resources.  
The following pairs can be determined similarly:
flows of goods and services -- prices; 
flows of trade of different kinds -- salary rates; 
flows of raw material resources -- rents; 
interest -- loan volume.

The asymptotics of $M$ and $\tilde{M}$ 
is given by the tunnel canonical operator
in the phase space of pairs.

We consider the phase space $R^{2n}$, where
the intensive variables play the role of coordinates 
and the extensive variables play the role of momenta.
In economics, the role of values of a random variable~$\la_i$ 
can be played by the prices of the corresponding goods, 
and~$N_i$ can, for example, be the number of sold goods, 
i.e., the number of people who bought goods of this particular
type, or the interest paid by the $i$th bank, etc.
Obviously, 
the price depends on the demand, i.e., 
$\la_i(N_i)$ is a curve in the two-dimensional phase space.
In the two-dimensional phase space,
to each point (vector) $\la_i$, $i=1, \ldots, n$,
there corresponds a vector $N_i (\la_1, \ldots, \la_n)$, 
$i=1, \ldots, n$.
In a more general case, this is an $N$-dimensional manifold 
(surface),
where the ``coordinates" and ``momenta" locally depend on~$n$
parameters, 
and the following condition is satisfied: 
the Lagrange brackets of the ``coordinates" and ``momenta"
are zero with respect to these parameters.
Hence such a manifold was called a Lagrangian manifold
by the author.
In other words, the form $\sum N_i d \la_i$ is closed 
(see the Afterword in~\cite{Heding} and \cite{TeorVozm65}). 
This means that $\int N_i d \la_i$ is independent of the path
and is called an action like $\int pdq$ 
($p$ is the momentum and $q$ is the coordinate) 
in mechanics.

The producer acquires resources and transforms 
the resources expenditure vector  
into the vector of material wealth production. 
Then the consumer acquires this material wealth. 
Thus, the equilibrium prices of resources and 
of the consumer material wealth 
are determined according to the above relations. 

In addition to such equilibrium prices,
there can also be vertical pairs for some types of material wealth 
and seller--buyer pairs 
(i.e., permanent seller -- permanent buyer pairs),
and the prices related to these pairs are formed.
This is an analog of the Cooper pairs in quantum statistics.

This construction requires the use of the ultra-secondary
quantization method in abstract algebraic form, 
which could be applied in economics.
The ``vertical'' clusters are also formed in this theory.

Thus, we have determined the parameter $\beta$, 
and the problem is solved completely.

{\bf{Example.}} 
In what follows, a person who buys stocks will be called a
player. 
Suppose that there are only two types of stocks,
the first are conditionally called  ``cheap'' stocks  
and the stocks of the second type 
are said to be ``expensive.''
We assume that a player buys a packet of $N$ stocks
in which the number of cheap stocks is~$N_1$
and the number of expensive stocks is $N_2=N-N_1$,
respectively. 
A player spends money to buy stocks, 
and the number of purchased stocks affects 
the price of stocks of both types.
In particular, the larger is the number of expensive stocks 
bought by a player, the less their price will be.
Consequently, in what follows, we assume that 
the player's expenditures for a packet of stocks 
depend nonlinearly 
on the number of purchased cheap and expensive stocks.
For example, it depends quadratically as follows: 
\begin{equation}
\cE(N_1)=\lambda_1N_1+\lambda_2N_2-\frac{\gamma N_1^2}{2N}-
\frac{\gamma N_2^2}{2N}=\lambda_2N-\frac{\gamma N}{2}
+(\lambda_1-\lambda_2+\gamma)N_1-\frac{\gamma N_1^2}{N},
\label{1d}
\end{equation}
where the numbers $\lambda_1$, $\lambda_2$, and $\gamma$
satisfy the conditions 
\begin{equation}
\frac{\gamma}{2}<\lambda_1<\lambda_2,\qquad
\lambda_2-\lambda_1<\gamma <2(\la_2 - \la_1). \label{2d}
\end{equation}
It follows from conditions~(\ref{2d}) that the function~(\ref{1d}) 
with $N_1=0,1,\dots,N$ has the global minimum for $N_1=N$ 
and a local minimum for $N_1=0$.

\begin{figure}[htbp]
\includegraphics[]{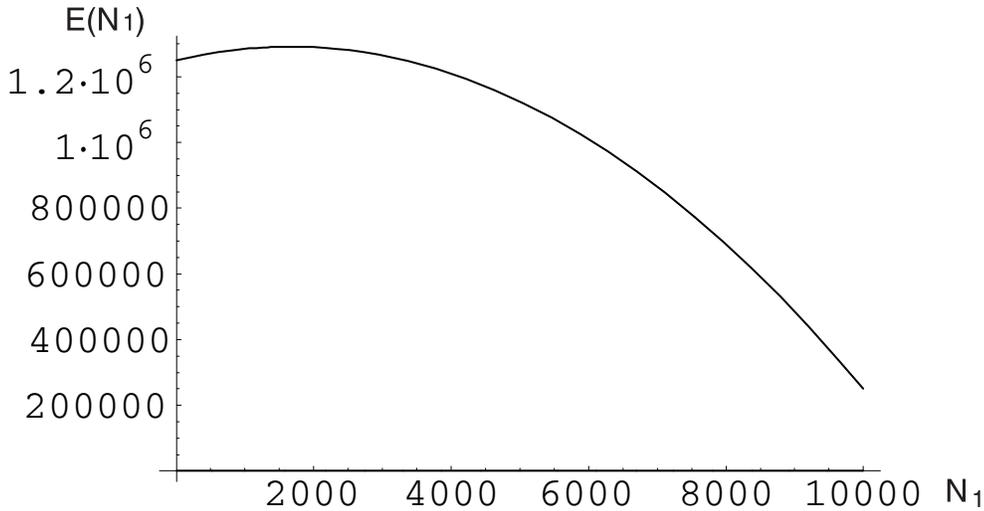}
\vskip-0.5cm
\caption{Graph of $E(N_1)$ for $T=0$}
\end{figure}

If, at the initial moment,  
the player buys~$N_1$ cheap stocks 
and $N_2$ expensive stocks so that 
$N_1 < \frac{\la_1-\la_2+\gamma}{\gamma} N$, 
then selling one cheap and one expensive stock 
so as to decrease $\cE(N_2)$, 
he will come to a local minimum at $N_1=0$,
i.e., he will buy all expensive stocks.
But if $N_1 > \frac{\la_1-\la_2+\gamma}{\gamma}N$,
then, as a result of a monotone process,  
the player will buy all cheap stocks.

Now we consider local financial averagings 
of the player income. 
We assume that $G_1$ dealers sell cheap stocks 
and $G_2$ dealers sell expensive stocks. 
In this case, the number of different ways 
in which the player can buy a packet of stocks is 
\begin{equation}
\Gamma(N_1)=\frac{(N_1+G_1-1)!}{(G_1-1)!N_1!}
\frac{(N-N_1+G_2-1)!}{(G_2-1)!(N-N_1)!}. \label{3d}
\end{equation}

{\bf Remark.} Instead of introducing different dealers,
we can assume that 
the cheapest and the most expensive stocks 
are, respectively, of $G_1$ and $G_2$ different types, 
but of the same price.

We assume that, for $\be= \infty$,
the player is at the point of local minimum for $N_1=0$. 
Since he tries to change the stocks pairwise and gradually 
(monotonically) so that not to increase his expenditures 
\footnote{The least risk principle in economics.}, 
we can consider the averaging 
only in a neighborhood of the point of local minimum
(the local financial averaging). 
If $\be$ varies slowly and $N\to \infty$, 
then the asymptotics of $M_\be$ as $N\to \infty$
again corresponds to the local minimum 
\begin{eqnarray}
&&\cE(N_1) = \be(\lambda_1N_1+\lambda_2N_2
-\frac{\gamma N_1^2}{2N}-\frac{\gamma N_2^2}{2N}) + \nn \\
&&+\ln \frac{(N_1+G_1-1)!}{(G_1-1)!N_1!}
\frac{(N-N_1+G_2-1)!}{(G_2-1)!(N-N_1)!}. \label{3dd}
\end{eqnarray}

\begin{figure}[htbp]
\includegraphics[width=10cm, height=18cm]{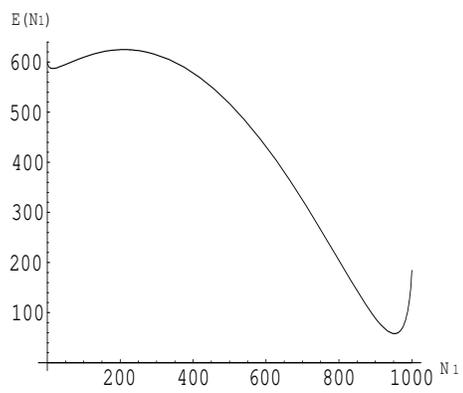} 
\vskip-0.5cm
\caption{Graph of $E(N_1)$ for $T=5$, $G_2=30$, and $\gamma=1.5$}
\end{figure}

Figure~3 shows how the entropy depends on the temperature and 
the local and global minima.

\begin{figure}[htbp]
\includegraphics[width=10cm, height=18cm]{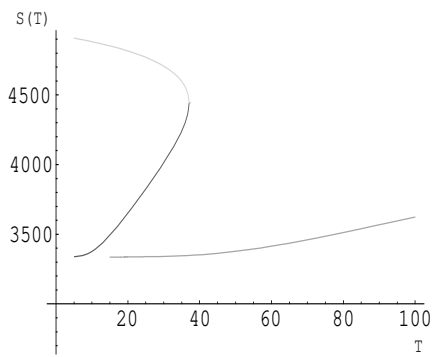} 
\vskip-0.5cm
\caption{The heavy line corresponds to the local minimum,
the global minimum lies below.
At the point $T=40$, the local minimum and the global maximum 
coincide and do not exist for $T>40$.}
\end{figure}

The curve breaks at the point $T\approx 40$.
At this point, 
the derivative $\frac{\pa S}{\pa T}$ becomes infinite, 
and the modified Laplace method, 
which could be used for the asymptotics 
of the local $M_\be$ at the other points,
cannot be used near this point.
The local minimum obtained by computer
shows unstable ``spreading."
It turns out that the asymptotics near this point 
can be expressed in terms of the Airy function
of an imaginary argument,
and this fact removes all the problems listed above.

For $T>40$, no equilibrium can exist 
when small changes in purchasing and selling occur. 
Therefore, to return to the equilibrium point
formed as a result of changes in the local averaging 
near another local minimum,
the player must change a large amount of stocks at once 
(see Fig.~3).

A similar situation occurs when $\ka=1$ 
and the player wants to win as much as possible.
Then the points of minima are replaced by the points of maxima 
between which, 
in the case of a quadratic dependence on $N_1$ and $N_2$, 
there is a minimum. 
If~$T$ varies from zero to some~$T_0$
at which the local maximum disappears,
then a jump occurs, which can be treated as 
a stock price break-down.
In our case, 
if the player simultaneously and very fast 
sells a large amount of stocks of one type 
and buys a large amount of stocks of a different type,
then he can again get to another equilibrium point. 

We have considered only the simplest model. 
In a more complicated economical situation
concerning the interests of a great mass of the population,
the people cannot change their behavior very fast,
passing, for example, 
from the usual consumer basket to a different basket
and thus changing their mode of life.
Then there is no equilibrium (balance) point in general,
and a sharp disbalance leads to general default.

{\bf{Generalization.}} 
In our example, we obtained a one-dimensional curve
corresponding to a local minimum of the entropy dependence
on the ``temperature" in the two-dimensional $S,T$-space
and considered its projection on the $T$-axis at the
point~$T_0$. 
This projection is not ``good," 
and we said that the asymptotics of~$M_\be$
in a neighborhood of this point must be replaced 
by the Airy function.

What picture appears in the general case 
where we have two ``conjugate" pairs: 
the entropy--temperature pair and, for example, 
the number of people~$N$ corresponding to 
some average salary~$\ve_k$. 

In this case, we consider a four-dimensional (phase) space,
where $T,\ve$ are the ``coordinates''
and $N,S$ are the momenta. 
The surface corresponding to our curve 
is two-dimensional and can (locally) be written 
in parametric form as 
\begin{equation}
T=T(\al_1, \al_2), \quad \ve = \ve(\al_1, \al_2),\quad 
N= N(\al_1, \al_2), \quad S=S(\al_1, \al_2), \nn
\end{equation}
where $\al_1, \al_2$ are parameters.

Because this surface must be obtained 
(at least, at the simple points of projection 
on the ``coordinate" plane)
from the asymptotics of sums of the form 
$$
M_\be = \frac {1}{\ka\be} \ln(\frac{e^{\ka\be a}+ e^{\ka\be
b}}{2}),\quad \ka =\pm 1, \quad  \be>0,
$$
it must be a Lagrangian manifold
(this notion was introduced by the author 
in \cite{TeorVozm65}). 
At the points of ``bad projection" on the $T,\ve$-plane,
the asymptotics is given by the tunnel canonical operator
\cite{TeorVozm65}, and its simplification,
depending on the form of the surface near 
the point of ``bad projection," 
can be obtained using~\cite{ArnoldVar}.

These general considerations can help constructing
the corresponding model of a given economical situation 
if the appropriate statistical data are available.

\section{The law of large numbers}

We consider the sellers who sell goods of $M$ types.
The corresponding {\it structure of sold goods}
is characterized by the vector
$\omega^N=\langle K_1,\dots, K_M\rangle$, 
where $K_i$ is the number of goods of the $i$th type, 
$i=1,\dots, M$, and 
$\sum\limits_{i=1}^M K_i\le N$. 
The {\it structure of purchased goods}
is considered similarly. 
Suppose that $\epsilon_i$ is the price of goods of the $i$th
type. 
The income obtained by realizing $K_1+\dots +K_n$ units of goods
is $E(\omega^N)=\sum\limits_{i=1}^M\epsilon_i K_i$.

The concepts introduced below 
are based on the notion of nonlinear averaging  
of the incomes obtained by realizing $\le N$ units of goods  
over all possible versions $\omega^N$  
of the structure of sold goods 
for a given positive value of the parameter~$\beta$:
\begin{eqnarray}
M_N(\beta)=\frac{1}{\beta}\log\left(\frac{1}{L}
\sum\limits_{\omega^N} 2^{\beta E(\omega^N)}\right)=
\frac{1}{\beta}\log\left(\frac{1}{L} \sum\limits_{K_1+\dots+K_M\le
N} 2^{\beta\sum\limits_{i=1}^n\epsilon_i K_i}\right),
\label{average-1a}
\end{eqnarray}
where $L$ is the number of terms in the sum. 
The nonlinear averaging is discussed in detail
in~\cite{Mas2003}. 
In the present paper, we show that,
under certain natural assumptions, 
the average can be calculated only in two ways: 
as the usual linear average 
and as the nonlinear average of the form (\ref{average-1a}).

We divide the goods of all types into groups 
with different prices.  
Suppose that the total number of groups of such goods is~$n$,
$n\le M$,
the $i$th group contains $G_i$ types of goods 
with the common price $\lambda_i$
and $\lambda_1<\lambda_2<\dots <\lambda_n$. 

In our statement, each $\lambda_i$ is equal to one of~$\epsilon_j$.
We denote 
$\overline{\lambda}=(\lambda_1,\lambda_2, \dots,\lambda_n)$. 
By definition, we have $\sum\limits_{i=1}^n G_i=M$.
Let $K_{i,j}$ be the total number of units of goods of the $j$th
type in the $i$th group of goods with the common price~$\lambda_i$. 
Thus, the structure of sold goods is given 
by the vector 
\begin{eqnarray}\label{stru-1}
\omega^N=\langle K_{1,1},\dots, K_{1,G_1},\dots, K_{i,1},\dots,
K_{i,G_i},\dots, K_{n,1},\dots, K_{n,G_n}\rangle.
\end{eqnarray}
Let $N_i=\sum\limits_{j=1}^{G_i}K_{i,j}$ 
be the number of all units of goods with the price $\lambda_i$, 
and let $\nu_i=N_i/N$ be the part of all such goods  
in their total amount, 
where $i=1,\dots, n$. 
The {\it price structure} of sold goods 
is characterized by the vector 
$\langle N_1,\dots, N_n \rangle$ 
or, in fractions, by the vector 
\begin{eqnarray}\label{stru-2}
\langle\nu_1,\dots, \nu_n\rangle.
\end{eqnarray}
Then the total cost of all $N$ units of goods is 
\begin{eqnarray}\label{income-1}
E(\omega^N)= \sum\limits_{i=1}^{n}\lambda_i
N_i=N\sum\limits_{i=1}^{n}\lambda_i\nu_i.
\end{eqnarray}
By $\Xi_N^{N_1,\dots, N_n}$ we denote the set of all versions 
of the structure of sold goods 
with a prescribed price structure 
$\langle N_1,\dots, N_n\rangle$. 
The number of elements in this set is\footnote{$|A|$ denotes 
the number of elements in the set~$A$. }
\begin{eqnarray}
|\Xi_N^{N_1,\dots, N_n}|= \binom{N_1+G_1-1}{G_1-1}
\cdot\dots\cdot \binom{N_n+G_n-1}{G_n-1}. \label{xi-1}
\end{eqnarray}
If the goods are separated in groups, 
then the sum (\ref{average-1a}) becomes 
\begin{eqnarray}
M_N(\beta,\overline{\lambda})=\frac{1}{\beta}\log\left(\frac{1}{L}
\sum\limits_{N_1+\dots+N_n\le N} |\Xi_N^{N_1,\dots, N_n}|
2^{\beta\sum\limits_{i=1}^n\lambda_i N_i}\right),
\label{average-1s}
\end{eqnarray}
where $L$ is the total number of all possible versions 
of the structure of sold goods, 
and the second sum in (\ref{average-1a})
is taken over all the price structures 
$\langle N_1,\dots, N_n\rangle$.

We assume that the sellers have sold at most $N$
units of goods of different types.
We consider the problem of predicting
the ``most expected'' structure of goods realized 
in the free market,
which is necessary for the average income
to be at least~$E_1$. 
Thus, the following inequality must hold:
\begin{eqnarray}\label{income-2}
\sum\limits_{i=1}^{n}\lambda_i N_i\ge E_1.
\end{eqnarray}
In addition, 
the following natural inequality must hold:
\begin{eqnarray}\label{income-2g}
\sum\limits_{i=1}^{n}N_i\le N.
\end{eqnarray}
We introduce the main definitions.
The base~$2$ logarithm of the number of possible 
versions of sale of~$N$ units of goods  
for which the income is no less than~$E_1$, 
i.e., (\ref{income-2}) is satisfied,
will be called the entropy $H(E_1,\overline{\lambda},N)$.

By $\theta(E_1,\overline{\lambda},N)$
we denote the solution (with respect to~$\theta$)
of the equation
\begin{eqnarray}
\frac{\partial
M_N(\theta^{-1},\overline{\lambda}))}{\partial\theta}=
H(E_1,\overline{\lambda},N). \label{def-te}
\end{eqnarray}

The solution of this equation exists and is unique. 
This follows from the monotonicity.

We set
\begin{eqnarray}
\tilde\nu_i(E_1,\overline{\lambda},N))
=\frac{1}{N} \frac{\partial M_N} {\partial\lambda_i}
|_{\theta=\theta(E_1,\lambda)}
\label{nu-i-1}
\end{eqnarray}
for $i=1,\dots, n$.

The notions introduced above are of general character and can be
used for all possible versions of the goods distribution.
In what follows, we consider a specific case 
for which we find the asymptotics of the introduced variables 
and prove a certain simplest version of the 
``law of large numbers'' under the following conditions.

We introduce the notation $\nu_i=N_i/N$ and $p_i(N)=G_i/N$ 
for $i=1,\dots,n$. 
We assume that the variables $p_i(N)$ have the limit $p_i$ 
as $N\to\infty$, and this limit is greater than zero 
for all~$i$. We also assume that $E_1=e_1 N$.
Let $\rho=N/M$ be a constant 
(one can also assume that $N/M\to\rho$ as $N\to\infty$).

In what follows, we also need the restriction
$e_1\le\rho\sum\limits_{i=1}^n\lambda_i p_i$ 
on the value of the average income per unit sold goods.

\begin{theorem}
Let $n\ge 3$.
Then, for an arbitrary $\epsilon>0$, 
the part of all versions 
of the structure of sold goods {\rm(\ref{stru-2})}
for which the average income per unit goods
is no less than $e_1$
\footnote{i.e., under restrictions {\rm(\ref{income-2})} 
and (\ref{income-2g}). } 
and $|\nu_i-\tilde \nu_i|\ge\epsilon$
at least for a single~$i$, $1\le i\le n$, 
does not exceed $2^{-c\epsilon^2N}$, 
where $c$ is a constant.
\end{theorem}

The conditions of the theorem contain 
the assumption that the number of nomenclatures of goods 
with the same price is sufficiently large.
But one can unite a group of goods with different prices 
and, calculating the nonlinear average of these prices, 
replace this group of prices in the statement of the theorem 
by this nonlinear average price.

We shall write the corresponding formula.

We consider a set of prices $\lambda_i$, 
where $i=1,\dots,n$, 
and a set of numbers $g_i$ equal, for example,
to the number of goods of different types but with the same
price $\lambda_i$. 
The numbers $g_i$ are quantities of the order of~$1$. 
By~$N_i$ we denote the number of goods purchased at the price
$\lambda_i$. 
Taking into account the fact that 
$g_i$ different goods can be purchased 
at the price $\lambda_i$,
the number of different way for buying $N_i$ goods 
at the price $\lambda_i$ is given by the formula
\begin{equation}
\gamma_i(N_i)=\frac{(N_i+g_i-1)!}{N_i!(g_i-1)!}, \label{1ge}
\end{equation}
and the number of different ways for buying 
the set of goods 
$\{N\}=N_1,\dots,N_n$ is, respectively, equal to  
\begin{equation}
\Gamma(\{N\})=\prod_{i=1}^n\gamma_i(N_i)=
\prod_{i=1}^n\frac{(N_i+g_i-1)!}{N_i!(g_i-1)!}. \label{2ge}
\end{equation}

We assume that the goods are divided in $m\le n$ 
groups as follows.
Suppose that there are two sequences~$i_\alpha$ 
and~$j_\alpha$, where $\alpha=1,\dots,m$,
such that 
\begin{eqnarray}
&&i_\alpha\le j_\alpha,\qquad
i_{\alpha+1}=j_\alpha+1,\qquad\alpha=1,\dots,m,
\nn\\
&&i_1=1,\qquad j_m=n. \label{3ge}
\end{eqnarray}
In this case, 
we say that the goods belong to the group of goods 
with number~$\alpha$ if their price $\lambda_i$ 
satisfies the condition $i_\alpha\le i\le j_\alpha$. 
By $\wt{N}_\alpha$ we denote the number of purchased goods
from the group with number~$\alpha$.
This number is given by the formula
\begin{equation}
\wt{N}_\alpha=\sum_{i=i_\alpha}^{j_\alpha}N_i. \label{4ge}
\end{equation}
We introduce $w_\alpha$, which is the nonlinearly averaged price
of goods in group~$\alpha$,
\begin{equation}
w_\alpha=\frac1{\beta\wt{N}_\alpha}\left(
\log\left(\sum_{\{N\}}^\alpha\prod_{i=i_\alpha}^{j_\alpha}\gamma_i(N_i)
2^{-\beta\lambda_iN_i}\right)
-\log\left(\wt{\gamma}_\alpha(\wt{N}_\alpha)\right)\right), 
\label{5ge}
\end{equation}
where $\beta$ is an economic parameter, 
for example, the volatility.
We also use the notation 
\begin{equation}
\sum_{\{N\}}^\alpha, \label{6ge}
\end{equation}
which means that the summation is performed 
over all the sets of nonnegative integers 
$N_{i_\alpha},\dots, N_{j_\alpha}$
such that 
\begin{equation}
\sum_{i=i_\alpha}^{j_\alpha}N_i=\wt{N}_\alpha. 
\label{7ge}
\end{equation}
Moreover, 
$\wt{\gamma}_\alpha(\wt{N}_\alpha)$ is given by the formula
\begin{equation}
\wt{\gamma}_\alpha(\wt{N}_\alpha)=\sum_{\{N\}}^\alpha
\prod_{i=i_\alpha}^{j_\alpha}\gamma_i(N_i). 
\label{8ge}
\end{equation}
We can show that~(\ref{8ge}) satisfies the formula
\begin{equation}
\wt{\gamma}_\alpha(\wt{N}_\alpha)
=\frac{(\wt{N}_\alpha+\wt{g}_\alpha-1)!}
{\wt{N}_\alpha!(\wt{g}_\alpha-1)!}, \label{9ge}
\end{equation}
where
\begin{equation}
\wt{g}_\alpha=\sum_{i=i_\alpha}^{j_\alpha}g_i. \label{10ge}
\end{equation}

We note that, in general,  
the nonlinearly averaged price of goods in group~$\alpha$ 
depends on~$\beta$ and~$\wt{N}_\alpha$.

Now we consider the nonlinear expectation of the income
\begin{equation}
M(\beta,\lambda,N)=-\frac1\beta\log\left(\sum_{\{N\}}'\prod_{i=1}^n\gamma_i(N_i)
2^{-\beta\lambda_iN_i}\right), \label{11ge}
\end{equation}
where
\begin{equation}
\sum_{\{N\}}' \label{12ge}
\end{equation}
denotes the sum over all sets of nonnegative integers
$\{N\}=N_1,\dots,N_n$
such that 
\begin{equation}
\sum_{i=1}^nN_i=N. \label{13ge}
\end{equation}

\begin{proposition}
The nonlinear expectation~{\rm(\ref{11ge})}
satisfies the relation
\begin{equation}
M(\beta,\lambda,N)=-\frac1\beta\log
\left(\sum_{\{\wt{N}\}}'\prod_{\alpha=1}^m
\wt{\gamma}_\alpha(\wt{N}_\alpha)2^{-\beta
w_\alpha\wt{N}_\alpha}\right), \label{14ge}
\end{equation}
where
\begin{equation}
\sum_{\{\wt{N}\}}' \label{15ge}
\end{equation}
denotes the sum over all sets of nonnegative integers 
$\{\wt{N}\}=\wt{N}_1,\dots,\wt{N}_m$
such that 
\begin{equation}
\sum_{\alpha=1}^m\wt{N}_\alpha=N. \label{16ge}
\end{equation}
\end{proposition}

This assertion readily follows from the definitions 
of nonlinearly averaged prices~(\ref{9ge}) 
and nonlinear expectation of income~(\ref{11ge}).

Now we have 
$$
\tilde \nu_\al = \frac{\pa M}{\pa \omega_\la},
$$
and the law of large numbers theorem remains the same, 
where $\tilde \nu_\al$ is the number of goods 
purchased at the prices $\la_i$ 
for $i_\al \leq i \leq j_\al$.

\end{document}